\documentclass[aps,prl,twocolumn,superscriptaddress,nofootinbib]{revtex4-2}

\usepackage{amsmath,amssymb}
\usepackage{graphicx}
\usepackage{bm} 
\usepackage{hyperref}
\usepackage{tikz}
\usetikzlibrary{arrows.meta} 

\definecolor{darkred}{rgb}{0.5,0.0,0.0}
\definecolor{darkblue}{rgb}{0.0,0.0,0.9}
\definecolor{darkgreen}{rgb}{0.0,0.5,0.0}

\newcommand{\hb}{\hbar}
\newcommand{\Seff}{ S_\text{eff}}

\newcommand{\im}{\operatorname{Im}}

\newcommand{\Ein}{\operatorname{Ein}}
\newcommand{\DV}{\Delta_V}
\newcommand{\DL}{\Delta_L}

\newcommand{\GR}{ {\Gamma_{\mathbb{R}}}}
\def \cJ {\mathcal{J}} 

\def \cO {\mathcal{O}}
\def \cZ {\mathcal{Z}}

\def \cI {\mathcal{I}}

\newcommand{\Tr}{\operatorname{Tr}}
\def \cP {\mathcal{P}}

\newcommand{\SmFigEight}{%
\begin{tikzpicture}[baseline=-0.5ex, scale=0.65]
  \draw[line width=0.6pt] (-0.3,0) circle (0.3cm);
  \draw[line width=0.6pt] (0.3,0) circle (0.3cm);
  \fill (0,0) circle (1.5pt);
\end{tikzpicture}}
\newcommand{\SmTheta}{%
\begin{tikzpicture}[baseline=-0.5ex, scale=0.65]
  \draw[line width=0.6pt] (0,0) circle (0.4cm);
  \fill (0,0.4) circle (1.5pt);
  \fill (0,-0.4) circle (1.5pt);
  \draw[line width=0.6pt] (0,0.4) -- (0,-0.4);
\end{tikzpicture}}
\newcommand{\SmDumbbell}{%
\begin{tikzpicture}[baseline=-0.5ex, scale=0.65]
  \fill (-0.45,0) circle (1.5pt);
  \fill (0.45,0) circle (1.5pt);
  \draw[line width=0.6pt] (-0.45,0) -- (0.45,0);
  \draw[line width=0.6pt] (-0.45,0.3) circle (0.3cm);
  \draw[line width=0.6pt] (0.45,0.3) circle (0.3cm);
\end{tikzpicture}}

\begin{document}

\title{
Beyond the Dilute Instanton Gas: \\Resurgence with Exact Saddles in the Double Well}
\author{Aur\'elien~Dersy}
\author{Matthew~D.~Schwartz}
\affiliation{Department of Physics, Harvard University, Cambridge, MA 02138, USA}

\date{\today}

\begin{abstract}
The path-integral approach to the double well has long been limited by the dilute instanton gas approximation. We show that if the finite Euclidean-time structure is taken seriously by using exact saddles, the dilute gas can be sidestepped, allowing the partition function and energy levels to be computed systematically. At each instanton order, the full resurgent structure---which saddles contribute, what asymptotic growth is expected and how ambiguities cancel---is encoded in a finite-dimensional Picard--Lefschetz contour integral over the quasi-zero modes with a clear geometric interpretation. Working at finite $T$ is essential: the dilute instanton gas can only access the ground-state splitting, whereas the exact finite-$T$ computation systematically produces the non-perturbative energy splittings for all excited states, including their full dependence on the level number. The key ingredients---Weierstrass elliptic functions for the saddles, Lam\'e operators for the fluctuations and Picard--Fuchs equations for the periods---form a coherent mathematical framework that both overlaps and complements that of Exact WKB.
\end{abstract}

\maketitle


Perturbation theory in quantum mechanics generically produces divergent asymptotic series whose factorial growth encodes non-perturbative physics~\cite{Bender:1969si,Brezin:1977gk}. The symmetric double well is the simplest arena in which the full resurgent structure can be worked out. Two complementary approaches address it: Exact WKB, which gives a complete algebraic solution through Voros symbols and connection formulas~\cite{Voros1983,DelabaereDillingerPham1993,DelabaerePham1999}, and the Euclidean path integral, which provides the physical picture through expansions about instanton saddle points. While Exact WKB has been developed into a powerful tool~\cite{Alvarez:2004qew,DunneUnsal2014WKB,Bucciotti:2023trp,Basar:2017hpr,Gu:2022sqc,vanSpaendonck:2023znn}, the path-integral side has mostly been organized around enhancing the dilute instanton gas (DIG) in Bogomolny~\cite{Bogomolny:1980ur} and Zinn-Justin~\cite{Zinn-Justin:1981qzi} (BZJ). In the DIG, multi-instanton configurations are built by sewing together well-separated one-instanton like $\tanh$ kinks on a Euclidean circle of circumference~$T$ and taking $T\to\infty$. A key ingredient of the path integral approach is a prescription to compute the path integral around the various instanton--anti-instanton saddles at next-to-leading order. Such integrals are divergent due to the presence of a quasi-zero mode, whose integration contour runs along a ridge of the action, with a decreasing direction towards the small instanton--anti-instanton separation regime. BZJ handle this mode by performing an analytic continuation of the coupling $g\to -g$, dropping sub-leading terms and continuing back. The BZJ prescription produces imaginary contributions that precisely cancel the Borel ambiguities of the perturbative series, providing compelling evidence for resurgence. Within that framework, Zinn-Justin and Jentschura~\cite{Zinn-Justin:2004vcw,Zinn-Justin:2004qzw} conjectured a generalized Bohr–Sommerfeld quantization condition and gave the multi-instanton trans-series of the ground-state energy to high order.

Despite these successes, the BZJ approach has fundamental limitations. First, taking $T\to\infty$ discards all finite-$T$ corrections, so this limit gives the same splitting for every level and only the ground-state energy can be extracted.  Second, the DIG ignores instanton interactions, only providing the leading contribution at each instanton number, with no systematic framework for computing subleading corrections. Instanton--anti-instanton interactions are only derived when sewing together $\tanh$ functions, merely leading to approximate solutions. Third, the BZJ prescription is \textit{ad hoc}---it produces the right answer but is not derived from a principled Picard--Lefschetz decomposition of the path integral. Several groups have worked towards bridging this last gap. Richard and Rouet~\cite{Richard:1980ei,Richard:1981gn} first classified the exact finite-$T$ periodic saddle points as Weierstrass elliptic functions on the curve defined by the quartic potential, a result revisited by Tanizaki and Koike~\cite{Tanizaki:2014xba} in the context of real-time Picard-Lefschetz theory and placed in the broader framework of algebraic integrable systems by Nekrasov~\cite{Nekrasov:2018pqq}. Behtash \textit{et al.}~\cite{Behtash:2018voa} recast the BZJ prescription in Lefschetz-thimble language, starting from an action parameterizing the dependence on the instanton--anti-instanton separation, analyzing its associated thimble structure and describing the cancellation of the imaginary parts.  While these works make progress, a few questions remain open: how to justify the thimble decomposition, how to systematically verify the trans-series, and how to actually compute the energies purely within the path integral. In this Letter we carry out that program. Working with the exact finite-$T$ saddles enables a clean Picard--Lefschetz decomposition, with no dilute-gas approximation, no ad hoc analytic continuations, and a transparent connection between thimble geometry and resurgent trans-series.


\bigskip

\textit{Exact saddles.} We consider the double well $V(x) = \tfrac{1}{8}(x^2-1)^2$ with wells at $x=\pm 1$. The Euclidean equations of motion $\ddot{x} = V'(x)$ admit a family of solutions parametrized by the conserved energy $\varepsilon = \tfrac{1}{2}\dot{x}^2 - V(x)$, with the four turning points defining an elliptic curve. The two independent half-periods of this curve are
\begin{equation}
  \omega_P(\varepsilon) = \frac{1}{2}\oint_P \frac{dx}{\sqrt{2\varepsilon{+}2V}}\,,\quad
  \omega_N(\varepsilon) = \frac{1}{2}\oint_N \frac{dx}{\sqrt{2\varepsilon{+}2V}}\,,
  \label{eq:periods}
\end{equation}
where the $P$-cycle encircles the classically allowed region within a single well ($\omega_P$ is purely imaginary) and the $N$-cycle encircles the barrier region ($\omega_N$ is real). The corresponding period integrals of the momentum $P_0(x) = \sqrt{2\varepsilon+2V(x)}$ are
\begin{equation}
  S_P^0(\varepsilon) = \frac{1}{2}\oint_P P_0\,dx\,,\quad
  S_N^0(\varepsilon) = \frac{1}{2}\oint_N P_0\,dx\,.
  \label{eq:SPN}
\end{equation}
For example, integrating between the turning points delimiting the barrier region we find
\begin{equation}
  S_N^0(\varepsilon) = \frac{\pi(1{+}8\varepsilon)}{4\sqrt{2}}\;
  {}_2F_1\!\left(\tfrac{1}{4},\tfrac{3}{4},2;\,1{+}8\varepsilon\right)\,,
  \label{eq:SN0_2F1}
\end{equation}
while the integral for the allowed region gives
\begin{equation}
    S_P^0(\varepsilon) =   - \pi i \varepsilon  \; {}_2 F_1\!\left( \tfrac{1}{4},
  \tfrac{3}{4}, 2 ; -8 \varepsilon \right) \,.
\end{equation}
$S_P^0$ and $S_N^0$ are the leading order WKB phases in the allowed and forbidden regions respectively. These integrals and the half-periods are related by $\omega_{P,N} = \partial_\varepsilon S_{P,N}^0$ and satisfy the Picard--Fuchs equation (for $P$ and $N$) $\varepsilon(1+8\varepsilon)(S^0)'' = -\tfrac{3}{2}\,S^0$. This equation allows one to express any $\partial_\varepsilon^k S^0$ (and as it turns out any higher order $S^m$) as a linear combination of $S^0$ and $(S^0)'$ with coefficients that are rational functions of $\varepsilon$. 

The general solution to the Euclidean equations of motion for the double well is parameterized by~\cite{Richard:1980ei}
\begin{equation}
  x_{k,k'}(t) = x_j\!\left[1 + \frac{6(x_j^2-1)}
    {1-3x_j^2+24\,\wp(t-t_0;\omega_P,\omega_N)}\right],
  \label{eq:xkk}
\end{equation}
where $x_j$ is a turning point, $\wp$ is the Weierstrass function, and $t_0$ is the time-translation zero mode. With this parametrization, the equation for the conserved energy becomes the Weierstrass equation $\dot{\wp}^2 = 4 \wp^3 - g_2 \wp - g_3$, with $g_2, g_3$ rational functions of $\varepsilon$. Looking for saddles with periodic boundary conditions imposes a quantization condition on $\varepsilon$ expressed using the half-periods as
\begin{equation}
  T = 2k\,\omega_N(\varepsilon) + 2k'\,\omega_P(\varepsilon)\,,
  \label{eq:quantcond}
\end{equation}
with integers $k \geq 1$, $0 \leq k' \leq k$. The instanton number is $n = 2k$ and $k'$ counts windings around the imaginary period such that the real saddles have $k'=0$. The on-shell action is the Legendre transform of the period integral:
\begin{equation}
  S_{k,k'} = 2k\,S_N^0(\varepsilon) + 2k'\,S_P^0(\varepsilon) - \varepsilon\,T\,,
  \label{eq:Skk} 
\end{equation}
where at large $T$ (small $\varepsilon$), $S_N^0 \to S_I = 2/3$ and
\begin{equation}
  S_{k,k'} \to 2k\,S_I - 16k\,e^{-i\pi k'/k}\,e^{-T/(2k)} + \cdots \,.
  \label{eq:Skk_largeT}
\end{equation}

With the exact zero-mode removed, the one-loop determinant around the $(k,k')$ saddle factorizes exactly into elliptic-curve data:
\begin{equation}
  \det\nolimits'\!\cO_{k,k'} = -\bigl(2k\,\partial_\varepsilon\omega_N + 2k'\,\partial_\varepsilon\omega_P\bigr)
  \bigl(2k\,S_N^0 + 2k'\,S_P^0\bigr)\,,
  \label{eq:detprime}
\end{equation}
valid at any finite $T$ with no approximation.

\bigskip

\textit{Path integral decomposition.} The Picard--Lefschetz decomposition of the partition function $Z = \Tr\,e^{-HT/\hb}$ is $Z = \sum_{k,k'} \eta_{k,k'}\,Z_{k,k'}$, where $Z_{k,k'}$ is the contribution from the steepest-descent thimble $\cJ_{k,k'}$ passing through the $(k,k')$ saddle and $\eta_{k,k'}$ is the intersection number of the associated steepest-ascent thimble with the real cycle~$\GR$. From Eq.~\eqref{eq:Skk_largeT}, $S_{k,k'} \to 2kS_I + \cdots$ at large $T$, so each $(k,k')$ saddle contributes an overall factor $e^{-nS_I/\hb}$ with instanton number $n = 2k$, regardless of~$k'$. Since the action functional is real on~$\GR$, gradient flow preserves real paths; so if $\GR$ intersected the unstable manifold of a complex saddle, the downward flow would remain real, while needing to converge to a complex saddle point---a contradiction. Hence $\eta_{k,k'} = 0$ for $k'\neq 0$ and only real saddles contribute to the Picard-Lefschetz decomposition of the partition function. Complex saddles still govern the Stokes structure indirectly (e.g. the $(1,1)$ saddle is the Stokes point of the $\cJ_{1,0}$ thimble) but do not enter the decomposition directly.

For the $n=0$ saddles (constant solution at $x=\pm1$) the partition function is 
\begin{equation}
Z_0 = Z_\text{SHO}\,e^{\DV T},
\label{Z0form}
\end{equation}
where $Z_\text{SHO} = \det \cO_0 = 1/(2\sinh\frac{T}{2})$ is the harmonic-oscillator partition function and $\DV T$ are the connected vacuum bubbles. These can be computed via Feynman diagrams with the periodic propagator $\Delta_P(t_1,t_2) = \cosh(T/2 - |t_1{-}t_2|)/(2\sinh\frac{T}{2})$. At order $\hbar$ there are three 2-loop graphs that sum to
\begin{multline}
  \DV T =
  \vcenter{\hbox{\SmFigEight}}
  \;+\; \vcenter{\hbox{\SmTheta}}
  \;+\; \vcenter{\hbox{\SmDumbbell}}
  \\
  = \frac{\hbar T}{4}\,\frac{\cosh T + 2}{\cosh T - 1}
  = \hbar T\!\left(\frac{1}{4} + \frac{3}{2}\,e^{-T} + \cdots\right).
  \label{eq:DVT}
\end{multline}
Expanding $Z_0$ in powers of $e^{-T}$
and matching to the spectral decomposition $Z_0=\sum_N \exp(-E_N T)$ gives
\begin{equation}
  E_N = \hbar\kappa 
  - \hbar^2\!\left[\tfrac{1}{16} + \tfrac{3}{4}\kappa^2\right]
  + \cO(\hbar^3),
  \label{eq:EN_2loop}
\end{equation}
with $\kappa = N{+}\frac{1}{2}$, in exact agreement with Bender--Wu~\cite{Bender:1969si}. The DIG, which discards $e^{-T}$ corrections, can only produce the $N$-independent constant $-\hbar^2/4$ for the ground state energy.

For the real $(k,0)$ saddles with
$n = 2k \ge 2$, the partition function factorizes as
\begin{equation}
  Z_{k,0} = \sqrt{\frac{nS_I}{2\pi\hb}}\,T\;
  (\det\nolimits_\perp \cO_{k,0})^{-1/2}\,
  e^{\DV T + \DL}\,
  e^{-nS_I/\hb}\,
  \cZ_{k,0}\,,
  \label{eq:Zn_structure}
\end{equation}
where $\sqrt{nS_I/(2\pi\hb)}\,T$ is the integral over the exact zero-mode along with the associated collective-coordinate Jacobian. The fluctuation determinant with the zero mode and $n{-}1$ quasi-zero modes removed is $\det_\perp\cO_{k,0}$, related to Eq.~\eqref{eq:detprime} by $\det_\perp = \det'/\prod_m\lambda_m$. $\DV$ is the same as in Eq.~\eqref{Z0form} coming from connected vacuum bubbles around the $n=0$ saddle, while $\DL$ gives corrections to the connected diagrams from the instanton background. $\cZ_{k,0}$ is the thimble integral over the $n{-}1$ quasi-zero modes including their Jacobians. At large~$T$, $\DV$, $\DL$, and $\det_\perp\cO_{k,0}/\det\cO_0$ are all real and $T$-independent
while $\cZ_{k,0}$ is a \textit{polynomial} of degree $n{-}1$ in~$T$. The entire resurgent structure is encoded in~$\cZ_{k,0}$ alone.


\begin{figure}[t]
  \centering
  \includegraphics[width=\columnwidth]{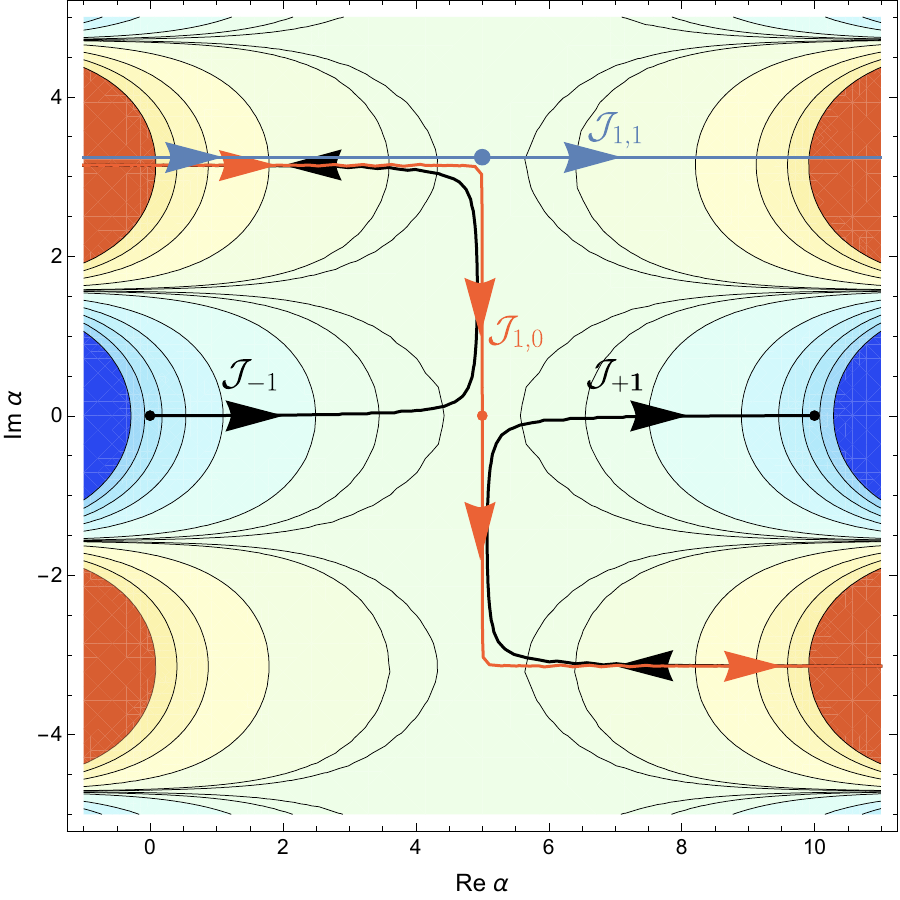}
  \caption{Real contours of the $n=2$ effective action~\eqref{eq:Seff2} with $\hb\to\hb\,e^{i\varepsilon}$. The real saddle (orange, $\alpha = T/2$) and complex saddle (blue, $\alpha = T/2 + i\pi$) are connected by a vertical segment. The thimbles $\cJ_{1,0}$ and $\cJ_{\pm1}$ can be expressed using three parts: the vertical arm from $\mp \infty \pm i\pi$ (giving $K_0$), the vertical segment (giving $I_0$), and the real line segment (encoding the Borel pole at $t=2S_I$).}
  \label{fig:thimbles}
\end{figure}

\bigskip

\textit{Instanton--anti-instanton contribution.} For the $n=2$ contribution, there is a single quasi-zero mode $\alpha$ (the instanton--anti-instanton separation). Near the real $(1,0)$ saddle, the effective action is~\cite{Bogomolny:1980ur,Zinn-Justin:1981qzi}
\begin{equation}
  \Seff(\alpha) = 2S_I - 12S_I\bigl[e^{-\alpha} + e^{-(T-\alpha)}\bigr]\,,
  \label{eq:Seff2}
\end{equation}
where the $12 S_I$ coefficient is determined by matching Eq.~\eqref{eq:Seff2} near its saddle to the large $T$ limit of the exact action (Eq.~\eqref{eq:Skk_largeT}) at $k'=0, k=1$. Contours of $\text{Re}(\Seff)$ and the thimble geometry are shown in Fig.~\ref{fig:thimbles}. This action increases from $\alpha=0$ to a maximum at $\alpha=T/2$ (the $(1,0)$ saddle) and then decreases back to $\alpha=T$. The complex saddle at $\alpha=T/2 + i\pi$ corresponds to the $(1,1)$ saddle of the full action. The real integration contour decomposes into $\GR=\cJ_{-1}+\cJ_{1,0} + \cJ_{1}$. The thimble integrals $\int_\cJ d\alpha\,e^{-\Seff/\hb}$ can all be decomposed into sums of 3 components. The \textit{vertical segment} between the real and complex saddles gives
\begin{align} 
  I_\text{vert}(T) &=  -i \pi\,e^{-2S_I/\hb}\,
  I_0\!\left(\frac{24S_I}{\hb}\,e^{-T/2}\right)\\
  &\xrightarrow{T\to\infty} -i \pi \,e^{-2S_I/\hb}\,,
  \label{eq:Ivert}
\end{align}
  where $I_0$ is the modified Bessel function. The \textit{arm} from $-\infty + i\pi$ to the complex saddle gives
\begin{align}
  I_\text{arm}(T) &= e^{-2S_I/\hb}\,
  K_0\!\left(\frac{24S_I}{\hb}\,e^{-T/2}\right) \notag\\
  &\xrightarrow{T\to\infty}
  e^{-2S_I/\hb}\!\left[\frac{T}{2} - \gamma_E
    - \ln\!\left(\frac{12S_I}{\hb}\right)\right].
  \label{eq:Iarm}
\end{align}
For the \textit{real segment} we start near the boundary at $\alpha_{\text{min}}=\ln 6$ (the point where $\Seff(\alpha)=0$) and integrate up to the real saddle. The result encodes the Borel singularity at $t=2S_I$:
\begin{align}
  I_\text{real}(T) &= e^{-2S_I/\hb}\!\left[
    -\Ein\!\left(-\frac{2S_I}{\hb}\right)
    + \frac{T}{2}  \right] + \cdots \,,
  \label{eq:Ireal}
\end{align}
where $\Ein(z) = \int_0^z dt\,(1{-}e^{-t})/t$ is the complementary exponential integral and the $\cdots$ represent finite corrections from the small $\alpha$ region that $\Seff$ does not model accurately. The thimble through the real $(1,0)$ saddle is $2I_\text{arm}+ 2I_\text{vert}$, giving $2 e^{-2S_I/\hb} K_0\!\left(-\tfrac{24S_I}{\hbar}e^{-T/2}+i\varepsilon\right)$, or at large $T$,
\begin{equation}
  \cZ_{1,0}
  = T - 2\gamma_E   - 2\ln\frac{12S_I}{\hb}\pm 2\pi i\,,
  \label{eq:J10full}
\end{equation}
with the sign of the imaginary part fixed by the small deformation $\hbar$ needed to disambiguate the Stokes phenomenon. The imaginary part is exactly canceled in the full path integral by the integral along the $\cJ_{-1}$ and $\cJ_{1}$ thimbles as can be seen clearly in Fig.~\ref{fig:thimbles}. 

The full thimble integrals, which sum to an integral along $\GR$, are the Borel resummation of the asymptotic series associated with the expansion about the various saddle points. In perturbation theory, we only see those series. For the $x=\pm 1$ saddles, the integral along $\alpha$ generates an asymptotic series which comes from the $I_\text{real}$ segment. Indeed, we have that $\Ein(-z) \sim \gamma_E + \ln(-z) - e^z \sum_{m\geq 0} m!/z^{m+1}$. Since $z=2S_I/\hbar$ this is an asymptotic series in $\hbar$ that would arise in perturbation theory from the vacuum bubbles in $\DV$. To appropriately include these contributions to the partition function, we have to use the same path integral measure in the $n=0$ and $n=2$ sectors, including both $\alpha$ and the collective coordinate $t_0$ whose integral leads to the extensive $T$ factor in the vacuum bubbles. Including both $x=\pm 1$ contributions and the Jacobian for $(t,\alpha)$ coordinates:
\begin{multline}
  Z = Z_\text{SHO}\frac{6S_I}{\pi\hb}\,T\Big[ 2\sum_m m!\left(\frac{\hbar}{2S_I}\right)^{m+1} \\
  + e^{-\frac{2S_I}{\hbar}}(T \pm2\pi i)+\cdots\Big]\,,
\end{multline}
where the $...$ indicate finite real subleading corrections. The imaginary part of the lateral Borel resummation of the $Z_0$ series exactly cancels the $2\pi i$ from $\cZ_{1,0}$, as guaranteed by the geometry.

\bigskip

\textit{Multi-instanton pairs contribution.} At $n=4$, the $(2,0)$ saddle has three quasi-zero modes. Two ``separation'' modes $x_1,x_2$ independently collapse adjacent instanton-anti-instanton pairs (connecting $n{=}4$ to $n{=}2$), while one ``breathing'' mode $y$ compresses all odd separations and expands all even ones (connecting
$n{=}4$ directly to $n{=}0$):
\begin{equation}
\label{eq:n4_mode_diagram}
\begin{tikzpicture}[xscale=0.1, yscale=0.5, baseline=(current bounding box.center),
  declare function={
    mytanh(\s) = (exp(\s)-exp(-\s))/(exp(\s)+exp(-\s));
  }]
  \draw[gray, dashed, thin] (-7,1) -- (67,1) node[right, black] {\footnotesize $+1$};
  \draw[gray, dashed, thin] (-7,-1) -- (67,-1) node[right, black] {\footnotesize $-1$};
  \draw[thick] (-7,-1) -- (-5,-1);
  \draw[thick, domain=-5:5, samples=60] plot (\x, {mytanh(0.5*\x)});
  \draw[thick] (5,1) -- (15,1);
  \draw[thick, domain=15:25, samples=60] plot (\x, {-mytanh(0.5*(\x-20))});
  \draw[thick] (25,-1) -- (35,-1);
  \draw[thick, domain=35:45, samples=60] plot (\x, {mytanh(0.5*(\x-40))});
  \draw[thick] (45,1) -- (55,1);
  \draw[thick, domain=55:65, samples=60] plot (\x, {-mytanh(0.5*(\x-60))});
  \draw[thick] (65,-1) -- (67,-1);
  \fill[orange] (0, 0) circle (4pt);
  \fill[blue!70] (20, 0) circle (4pt);
  \fill[orange] (40, 0) circle (4pt);
  \fill[blue!70] (60, 0) circle (4pt);
  \draw[<->, gray!80] (0, -1.6) -- node[below] {\footnotesize $\alpha_1$} (20, -1.6);
  \draw[<->, gray!80] (20, -1.6) -- node[below] {\footnotesize $\alpha_2$} (40, -1.6);
  \draw[<->, gray!80] (40, -1.6) -- node[below] {\footnotesize $\alpha_3$} (60, -1.6);
  \definecolor{modex1}{rgb}{0.78, 0.24, 0.20}
  \definecolor{modex2}{rgb}{0.18, 0.42, 0.80}
  \definecolor{modey}{rgb}{0.20, 0.58, 0.30}
  \draw[-{Stealth[length=5pt]}, thick, modex1] (1, 0.2) -- (6, 0.2)
    node[above, midway] {\footnotesize $x_1$};
  \draw[{Stealth[length=5pt]}-, thick, modex1] (14, 0.2) -- (19, 0.2)
    node[above, midway] {\footnotesize $x_1$};
  \draw[-{Stealth[length=5pt]}, thick, modex2] (21, 0.2) -- (26, 0.2)
    node[above, midway] {\footnotesize $x_2$};
  \draw[{Stealth[length=5pt]}-, thick, modex2] (34, 0.2) -- (39, 0.2)
    node[above, midway] {\footnotesize $x_2$};
  \draw[-{Stealth[length=5pt]}, thick, modey] (1, -0.2) -- (6, -0.2)
    node[below, midway] {\footnotesize $y$};
  \draw[{Stealth[length=5pt]}-, thick, modey] (14, -0.2) -- (19, -0.2)
    node[below, midway] {\footnotesize $y$};
  \draw[-{Stealth[length=5pt]}, thick, modey] (41, -0.2) -- (46, -0.2)
    node[below, midway] {\footnotesize $y$};
  \draw[{Stealth[length=5pt]}-, thick, modey] (54, -0.2) -- (59, -0.2)
    node[below, midway] {\footnotesize $y$};
\end{tikzpicture}
\end{equation}
Near the (2,0) saddle, dropping the cross-coupling between modes, the effective action is 
\begin{align}
  \Seff \nonumber
  =& 4 S_I -24S_I e^{-T/4}\bigl[\left(\cosh x_1-1\right) +\left(\cosh x_2-1\right) \bigr] \\&- 48S_I e^{-T/4} \left(\cosh y-1\right) 
  +\cdots\,.
  \label{eq:eq:Seff4}
\end{align}
At large $T$ the three-dimensional thimble integral factorizes in the normal-mode basis into a product of modified Bessel functions $\cI_{2,0} \sim  e^{-\frac{4S_I}{\hb}} [K_0(-z_4 + i\varepsilon)]^2[K_0(-2z_4 + i\varepsilon)]$ with $z_4 = 24S_I e^{-T/4}/\hb$, giving
\begin{multline}
  \cZ_{2,0} \sim \bigl[\tfrac{T}{2} - 2\gamma_E - 2\ln(12S_I/\hb) - 2\pi i\bigr]^2  \\  \times \bigl[\tfrac{T}{2} - 2\gamma_E - 2\ln(24S_I/\hb) - 2\pi i\bigr]\,.
  \label{eq:n4_largeT}
\end{multline}
The $2{:}1$ channel structure (two separation modes vs.\ one breathing mode) determines the three-way ambiguity cancellation structure
\begin{equation}
  \tfrac{1}{2}\im[\Delta_4 Z_0]
  + \tfrac{1}{2} \im[\Delta_2 Z_2]
  + \im[Z_4] = 0 \,,
  \label{eq:3wayZ}
\end{equation}
where $\Delta_n Z_m$ denotes the alien derivative~\cite{Dorigoni:2014hea}, i.e.\ the discontinuity of the lateral Borel resummation of $Z_m$ across the singularity at $t = nS_I$. This equation has an energy trans-series avatar $\tfrac{1}{2}\im[\Delta_4 E_0] +\tfrac{1}{2} \im[\Delta_2 E_2] + \im[E_4] = 0$ which we have verified explicitly using Exact WKB.


\bigskip

\textit{Energy spectrum.} Next, let us turn to the non-perturbative splittings of the energy levels. These are simplest to extract from the twisted partition function $\widetilde{Z} = \Tr(\cP\,e^{-HT/\hb})$ where $\cP$ is the parity operator. Since states split by parity, we anticipate $E_N^\pm = E_N \pm \exp(-S_I/\hb)\Delta_N$ so that to leading order in $e^{-S_I/\hb}$, after using $E_N=(N+1/2)\hbar + \cdots$,  the splitting is
\begin{equation}
\widetilde{Z} \approx \sum_N - e^{-\frac{S_I}{\hb}} e^{-\frac{E_N T}{\hb}}\frac{2T\Delta_N}{\hb}
=-e^{-\frac{S_I}{\hb}}  \frac{2T}{\hb} e^{-T/2} g(u)
\label{eq:Ztilde_expansion}
\end{equation}
where $g(u)=\sum_N \Delta_N u^N $ with $u=e^{-T}$.  The path integral corresponding to the twisted partition function has anti-periodic boundary conditions. Its saddle points are the same elliptic functions as in Eq.~\eqref{eq:xkk} but with a half-period quantization condition $T = (2k+1)\omega_N(\varepsilon) + 2k'\omega_P(\varepsilon)$. The least-action saddle is the one-instanton saddle which can be written in Jacobi elliptic form
\begin{equation}
  x_\cI(t) = \sqrt{\frac{2\sigma^2}{1{+}\sigma^2}}\;
  \operatorname{sn}\!\left(\frac{t}{\sqrt{2(1{+}\sigma^2)}},\,\sigma^2\right),
  \label{eq:n1_jacobi}
\end{equation}
where $\sigma^2 = (1{-}\sqrt{-8\varepsilon})/(1{+}\sqrt{-8\varepsilon})$. The one-loop fluctuation operator around a generic saddle $x_*$ is
\begin{equation}
  \cO = -\partial_t^2 + V''(x_*) = -\partial_t^2 + \tfrac{1}{2}(3x_*^2 - 1)\,.
  \label{eq:fluctop}
\end{equation}
As $T\to \infty$, the instanton $x_\cI$ approaches the DIG form $x_* = \tanh(t/2)$ and $\cO$ reduces to the P\"oschl--Teller operator $\cO_\text{PT} = -\partial_t^2 + 1 - \tfrac{3}{2}\operatorname{sech}^2(t/2)$. Around the exact Jacobi saddle~\eqref{eq:n1_jacobi}, after using $s = t/\sqrt{2(1+\sigma^2)}$, we instead find
\begin{equation}
  \cO_\cI = -\partial_s^2 + 6 \sigma^2\,\operatorname{sn}^2\!\!\left(s,\,\sigma^2\right),
  \label{eq:Lame}
\end{equation}
which is a Lam\'e equation. Both operators are exactly solvable. Both have a bound state and an infinite tower of positive modes, but the P\"oschl--Teller spectrum has a quasi-zero mode with $\lambda_0\sim e^{-T}$ while the Lam\'e spectrum has an exact zero mode at finite $T$.

The exact instanton action is $S[x_\cI] = S_N^0(\varepsilon) - \varepsilon T$ which, after using Eq.~\eqref{eq:SN0_2F1} and expanding the quantization condition to find $\varepsilon(T)$, expands at large $T$ as
\begin{equation}
  S[x_\cI] = S_I - 8\,e^{-T} + (136 - 48T)\,e^{-2T} + \cdots \,.
  \label{eq:S1_exact}
\end{equation}
 Combining the exact action, the collective-coordinate integral and Jacobian $T\sqrt{S_N^0/(2\pi\hb)}$, and the determinant in Eq.~\eqref{eq:detprime} at $2k=1$, we obtain
 for the one-instanton sector of the twisted partition function
\begin{equation}
  \widetilde{Z}_1(T) = \frac{T}{\sqrt{2\pi\hb\,\partial_\varepsilon\omega_N(\varepsilon)}}\,
  e^{-S[x_\cI]/\hb}\,,
  \label{eq:Ztilde1}
\end{equation}
where $\partial_\varepsilon\omega_N$ can be related to $S_N^0$ by the Picard--Fuchs equation.  Eq.~\eqref{eq:Ztilde1} is exact at one-loop and finite $T$. Expanding at large $T$ and matching to Eq.~\eqref{eq:Ztilde_expansion} gives
\begin{multline}
  g(u) = -2\sqrt{\frac{\hb}{\pi}}\,
  [1 - 46u - 12u\ln u + \cdots] \\
  \times e^{(8u - 136u^2 - 48u^2\ln u + \cdots)/\hb}\,.
  \label{eq:gu}
\end{multline}
Since $g(u)=\sum \Delta_N u^N$, the leading contribution $e^{8u/\hb}=\sum_m (8 u/\hbar)^m / m!$ immediately yields the leading non-perturbative splitting for all levels. For $\kappa=N+1/2$ we get
\begin{equation}
  \Delta_N = -\frac{\hb}{\sqrt{2\pi}}\,
  \frac{(8/\hb)^\kappa}{\Gamma(\kappa{+}\tfrac{1}{2})}\,.
  \label{eq:DeltaN}
\end{equation}
This splitting agrees with the result from Exact WKB, which requires either resumming the Voros symbol to all orders in the Riccati/WKB expansion~\cite{Voros1983,DelabaerePham1999,Sueishi:2020rug,Bucciotti:2023trp} or using a connection-formula approach~\cite{Alvarez:2004qew,Zinn-Justin:2004vcw,Zinn-Justin:2004qzw}. From the path integral, we can just directly read off the leading $T$ dependence in the exact action. 

Recall that in the DIG path integral~\cite{Coleman:1985rnk,Zinn-Justin:1981qzi}, the $T\to\infty$ limit is taken before extracting energies. There, the equivalent result is $\widetilde{Z}_1 \propto KT\,e^{-S_I/\hb}$, where $K$ is the usual one-loop DIG prefactor from the collective-coordinate Jacobian and the Pöschl--Teller determinant, which yields $\Delta_N = -K\hb\,e^{-S_I/\hb}$, a uniform splitting for each level. It is uniform because the $e^{-T}$ corrections to the action have been discarded. If we try to keep the corrections using the DIG action $S[\tanh(t/2)] = S_I - 4e^{-T} + \cdots$ the splitting would come out with a $(4/\hbar)^N$ factor instead of $(8/\hbar)^N$ which is incorrect. This mismatch originates from the hard truncation of the instanton profile in the overlap region with the boundary. To get all the factors right, one must use the exact finite $T$ instanton solution.

To proceed to the 2-loop non-perturbative correction, we must contend with the $\ln u$ terms in Eq.~\eqref{eq:gu}. These cannot match to the form $g(u) = \sum \Delta_N u^N$. These apparent singularities are resolved by also including the corrections to the perturbative energies in the spectral sum. Using the 2-loop energies in Eq.~\eqref{eq:EN_2loop}, expanding $\exp(-E_N T/\hb)$ perturbatively around $\exp(-\kappa T)$ gives additional $T = - \ln u$ terms. Then the spectral decomposition reads
\begin{equation} \label{eq:gu2l}
  g(u) = \sum_{N=0}^\infty \Delta_N\,u^N\, 
  \bigl[1 - \hb\,\tfrac{3N}{2}\ln u - \hb\,\tfrac{3N(N{-}1)}{4}\ln u + \cdots\bigr].
\end{equation}
This is compared to the expansion of Eq.~\eqref{eq:gu}:
\begin{equation}
  \label{eq:gu2m}
    g(u)=\sum_{N=0}^\infty \Delta_N u^N (1-46 u - 12 u \ln u  -  \tfrac{136 u^2}{\hbar}-\tfrac{48 u^2}{\hbar}\ln u)
\end{equation}
Using $\Delta_{N-1}/\Delta_N = N\hbar/8$ to convert the $N$-independent coefficients in Eq.~\eqref{eq:gu2m} to the $N$-dependent ones in Eq.~\eqref{eq:gu2l}, the $\ln u$ terms in the two expressions are seen to agree exactly. Remarkably, the $12u\ln u$ term comes from the one-loop functional determinant $\det'\cO_\cI$, the $48u^2\ln u$ comes from the tree-level action $S[x_\cI]$, and both agree with the spectral $\ln u$ that comes from the two-loop vacuum bubbles~\eqref{eq:DVT}. This three-way cancellation from three different loop orders is a stringent consistency check on the exact finite-$T$ computation, involving data from three different sources that are superficially unrelated.

Computing the $\cO(\hb^2)$ non-perturbative corrections from the path integral requires evaluating the two-loop vacuum bubbles in the instanton background. This is significantly harder than the perturbative calculation because the instanton propagator breaks time-translation invariance. Although a direct diagrammatic calculation has not yet been performed, the sophisticated mathematical structure surrounding the exact solutions---the Lam\'e spectrum of the fluctuation operator, the Picard--Fuchs differential equation relating periods and actions, and the closed-form elliptic-curve data entering the determinant~\eqref{eq:detprime}---suggests that an exact two-loop result may be within reach.

In summary, the central result of this Letter is that the path integral cleanly separates resurgence from spectral extraction, but this {\it only} works if exact results are kept at finite $T$, which is not possible within the dilute instanton gas.  All non-perturbative dynamics---which saddles contribute, what asymptotic growth is expected and how ambiguities cancel---is captured by finite-dimensional Picard--Lefschetz integrals over the quasi-zero modes, with a transparent geometric interpretation. This approach cleanly upgrades the ad-hoc BZJ analytic continuation prescription and the dilute instanton gas approximation. After translating to the energy spectrum, the clean resurgent structure gets lost in a web of contributions from different orders in perturbation theory.

Looking forward, the geometric understanding of resurgence developed here, in which ambiguity cancellation follows directly from Picard--Lefschetz theory, may extend to quantum field theories where the interplay between perturbative and non-perturbative effects is even richer and less well understood. In QCD~\cite{tHooft:1976snw} the dilute instanton gas is similarly unreliable. Its most visible failure, the divergence of the integral over the instanton size, is related to renormalon effects~\cite{tHooft:1977xjm,Beneke:1998ui,Bhattacharya:2024hhh} and isn't a problem of the instanton expansion itself. The genuine instanton difficulty is the treatment of multi-instanton states. Just as in the double well, exact multi instantons--anti-instantons saddles do not exist at infinite time/volume in QCD, and dilute-gas approximate solutions have an unstable attraction. This decay under gradient flow can be handled by an ad hoc valley prescription~\cite{Balitsky:1985in} with analytic continuation similar to BZJ. By analogy with the double well, one expects that compactification on $\mathbb{R}^3 \times S^1$ or $T^4$ should replace these approximate configurations with genuine critical points~\cite{Kraan:1998pm,Lee:1998bb,Argyres:2012ka} and make the Picard--Lefschetz decomposition rigorous. Although the calculations are much more difficult, it is reasonable to expect that a rigorous resurgent structure can emerge from the QCD path integral, provided that exact finite-volume saddles are used.

\begin{acknowledgments}
\textit{Acknowledgements.}~We thank Arindam Bhattacharya, Jordan Cotler, and Gerald Dunne for valuable conversations and G\"ok\c{c}e Ba\c{s}ar and Marco Serone for help with Exact WKB. Parts of this manuscript were written with the help of Claude Opus~4.6. AD and MDS take full responsibility for the correctness of the work. This work was supported by the U.S.\ Department of Energy, Office of Science, under grant DE-SC0013607.
\end{acknowledgments}

\bibliography{refs}

\end{document}